\documentclass[scriptaddress,onecolumn, prd,showkeys]{revtex4}

	\usepackage{amsmath}
	\usepackage{makeidx}
	\usepackage{amsfonts}
	\usepackage[ansinew]{inputenc}
	\usepackage[usenames,dvipsnames]{pstricks}
	\usepackage{subfigure}
	\usepackage{epsfig}
	\usepackage{pst-grad} 
	\usepackage{pst-plot} 
	\usepackage[colorlinks,hyperindex]{hyperref}
	\hypersetup
	{
		colorlinks,%
		citecolor=black,%
		linkcolor=black,%
		urlcolor=black,%
	}

	\newcommand{\ket}[1]{\left| #1 \right\rangle}
	\newcommand{\bra}[1]{\left\langle #1 \right|}

\makeindex

\begin{document}

\title{Stochastic quasi-classical wavefunction of the Universe from the third quantization procedure}

\author{P. Ivanov }
\email{pbi20@cam.ac.uk }
\affiliation{ Astro Space Centre, P. N. Lebedev Physical Institute, 84/32 Profsoyuznaya Street, Moscow, 117997, Russia \\
DAMTP, University of Cambridge, Wilberforce Road, Cambridge CB3 0WA, UK \\
}

\author{S. V. Chernov}
\email{ chernov@td.lpi.ru }
\affiliation{ Astro Space Centre, P. N. Lebedev Physical Institute, 84/32 Profsoyuznaya Street, Moscow, 117997, Russia \\ }

\date{today}

\begin{abstract}
We study quantized solutions of the Wheeler de Witt (WdW) equation describing a closed Friedmann-Robertson-Walker universe with a $\Lambda $ term and a set of
massless scalar fields. We show that when $\Lambda \ll 1$ in the natural units and
the standard $in$-vacuum state is considered, either wavefunction of the universe, $\Psi$,  or
its derivative with respect to the scale factor, $a$, behave as random quasi-classical fields at sufficiently large values of $a$. The former case is realised when $1 \ll a \ll e^{{2\over 3\Lambda}}$, while the latter is valid when $a \gg  e^{{2\over 3\Lambda}}$. The statistical r.m.s value of the wavefunction is proportional to the Hartle-Hawking wavefunction for a closed universe with a $\Lambda $ term. Alternatively, the behaviour of our system at large values of $a$ can be described in terms of a density matrix corresponding to a mixed state, which is directly determined by statistical properties of $\Psi$. Similar to  $\Psi$, the density matrix can be considered as c-number valued in the position and momentum representations. The probability distribution to find a universe with particular values of the scale factor and field amplitudes following from this density matrix is again proportional to that of the Hartle-Hawking wavefunction, while the probability distribution over field velocities is non-trivial and different from what follows from the Hartle and Hawking formalism.

We suppose that the same behaviour of $\Psi$ can be found in all models exhibiting copious production of excitations with respect to the $out$-vacuum state associated with classical trajectories at large values of $a$. Thus, the third quantization procedure may provide a 'boundary condition' for classical solutions of the WdW equation. Contrary to the previous proposals, in our case two equivalent descriptions of this classical solutions are possible. Either
$\Psi $ can be regarded as a stochastic classical quantity or the system can be viewed as being in
a mixed state defined over classical solutions to WdW equation.
\end{abstract}


\maketitle

\section{Introduction}
One of the possible approaches to the problem of quantum gravity is based on the canonical quantization of Einstein equations and analysis of properties and solutions of the emerging Wheeler de Witt (WdW) equation for the wave functional of the Universe, $\Psi$ \cite{dewitt},
see e.g. \cite{kif} for a review and further references. When $\Psi $
is considered as a c-number value some long standing issues arise. First, a direct probabilistic interpretation is hampered, second, there is an ambiguity related to initial (or, boundary) conditions for solutions to WdW equation. Over the years several proposals for a possible
choice of initial conditions have been made, notably the ones leading to the tunneling
wavefunction of Vilenkin \cite{vil} and 'no-boundary' wavefunction of Hartle and Hawking \cite{hh}.

Some problems are alleviated in an approach, where $\Psi$ itself is treated as operator valued, $\Psi \rightarrow \hat \Psi$. This is based on observation that the WdW equation has a formal structure of a hyperbolic equation with a variable determined by the volume element of spatial hypersurfaces playing a role of a 'time', which can again be canonically quantized. This procedure is called 'the third quantization formalism', see e.g. \cite{banks}. It was explicitly realised in minisuperspace models, where the metric was fixed by the condition that it is of Friedmann-Robertson-Walker type, and matter degrees of freedom were modeled as a set of scalar fields, see e.g. \cite{rub}, \cite{hm}.

In the minisuperspace approach the wave functional becomes a function depending on the scale factor, $a$, and field amplitudes, and WdW equation is reduced to a Klein-Gordon equation with second derivatives over the scale factor and the amplitudes entering it with opposite signs and, a potential being, in general, a function of these variables. When the logarithm of the scale factor is chosen as a time variable the differential part of WdW equation takes the standard form of the
d'Alembert operator in a certain factor ordering scheme assumed from now on. Close to singularity this variable tends to minus infinity and it may be shown that the potential tends to zero. In this limit the quantized WdW equation is the familiar Klein-Gordon equation for a massless
quantum scalar field, and, therefore, the standard vacuum state for this field can be chosen as a quantum state of the system, e.g. \cite{hm}. Hereafter, we call it the $in$-state having in mind the analogy with formalism of quantum fields in curved spacetimes, see \cite{bd}.
On the other hand, in a class of models, where classical dynamics with large values of scale factors is possible, there is another natural vacuum state associated with a set of quasi-classical solutions of WdW equation called hereafter the $out$-state. In general, this state does not coincide with the $in$-state, thus the latter contains excitations with respect to
the former one. This effect is interpreted as 'creation of universes from nothing' in the framework of the third quantization procedure \footnote{A care should be taken with this terminology. Rather, as in the general case of creation of particles
by an external field, see e.g. \cite{bd}, we are talking about amplification of vacuum fluctuations of the quantized wavefunction by evolving gravitational field.}.
Technically, this means that a positive frequency solution defined with respect to the $in$-state is a mixture of positive and negative. Care should be taken with this terminology. Rather, as in the general case of creation of particles by an external field, see e.g. \cite{bd}, we are talking about amplification of vacuum fluctuations of the quantized wavefunction by evolving with the scale factor gravitational field.frequency solutions corresponding to the $out$-state, with the Bogolyubov coefficient $\beta$ in the decomposition of positive frequency solutions over the negative frequency ones being nonzero. The amount of produced universes with given constants of motion corresponding to a particular quasi-classical $out$ mode is proportional to the square of absolute value of the $\beta$, summed over all modes defined with respect to the $in$-state.

In this Paper we would like to propose another interpretation of the emergence of 'classical' properties of the system on hand in the framework
of the third quantization formalism. Namely, we show that in a range of sufficiently large scale factors the quantized wavefunction can behave as a random classical variable.
We consider a simple WdW minisuperspace model of a closed FRW universe with a $\Lambda $ term and $n$ homogeneous massless scalar fields, $\varphi_i$, $2\le i \le n$. It was shown that when $\Lambda \ll 1$ in the natural Planck units, typical values of $\beta $ are exponentially
large, $\beta \propto e^{{2\over \Lambda}}$ \cite{rub}. On the other hand, the WdW equation of this model is quite similar to those used to describe the dynamics of test quantum fields in a natural $in$-vacuum state in inflationary models, see e.g. \cite{star}, \cite{l1}, \cite{lindeb}. As is known the latter problem also exhibits a copious 'particle' production with respect to a suitably chosen $out$-state. Also, a large scale part of these fields essentially behaves as a random classical field, e.g. \cite{star}, \cite{pol}, \cite{kifpol}.  In this case the creation and annihilation operators in the decomposition of the test field over the normal
modes can be treated as classical random variables obeying Gaussian statistics. There are several possible tests of this (quasi)classical behaviour, in particular based on investigation of the evolution of the $in$-vacuum state in the Schrodinger representation (e.g. \cite{gr2}, \cite{pol}), or the use of the Wigner function, e.g. \cite{hal}, \cite{pol}. In particular, the $in$-vacuum should be strongly squeezed with dispersion of a canonical variable manifesting
a quasi-classical behaviour being much larger than a typical one following from the uncertainty principle (e.g. \cite{gr2}). The Wigner function in the regime of quasi-classical dynamics is approximately reduced to a phase density distribution of a bunch of classical trajectories.

We show that both these criteria are satisfied in our model in a certain average sense, for modes giving the main contribution to the expectation values of interest \footnote{The fact that the vacuum state is strongly squeezed in a similar model was also note by \cite{sid}.}. Therefore,
these expectation values can be treated as statistical averages of classical quantities. However, an important difference with the test field case consists in the fact that in our case mode amplitudes, and, accordingly, $\hat \Psi$, are approximately quasi-classical only when $1 \ll a \ll e^{{2\over 3\Lambda}}$. When
$a \gg e^{{2\over 3\Lambda}}$ the mode momenta are quasi-classical, and, therefore, it is the derivative of  $\hat \Psi$ over $a$, which exhibits quasi-classical behaviour. Thus, the third quantization procedure together with a natural choice of the $in$-state may be used to define an initial condition for a classical wavefunction $\Psi$. Unlike the known proposals for the initial conditions in our case $\Psi $ is essentially a random quantity. Interestingly,
the rms value of our wavefunction, $\sqrt{<\Psi \Psi^*>}$, is approximately proportional to the Hartle and Hawking expression in our model, although it is not clear to us, whether this fact is due to a coincidence, or it may be of a generic nature.

Alternatively, when  $1 \ll a \ll e^{{2\over 3\Lambda}}$ we can use a density matrix with c-numbered matrix elements instead of the classical wavefunction. Thus, in our model
in this regime the quantum state of the Universe is a mixed one. Its diagonal elements in the position representation are equal to  ${<\Psi \Psi^*>}$, while it has a non-trivial structure in the momentum representation giving a probability to find universes with different values of $\dot \varphi_i$.

The structure of the Paper is as follows. In Section  \ref{basic} we introduce basic definitions and equations. In Section \ref{wdw} we obtain asymptotic solutions to the WdW equation in the limit $\Lambda \rightarrow 0$ based on a procedure involving the Wentzel-Kramers-Brillouin-Jeffreys (WKBJ) technique and the use of different solutions with common ranges of validity. In Section \ref{trans} we discuss values of $|\beta|^2$ and the behaviour of the Wigner function and the vacuum state in our model, showing that they all indicate quasi-classical dynamics of the variables of interest in the sense explained above. In Section \ref{wave} an explicit form of $\Psi$ valid for  values of $a \gg 1$ is obtained and its averaged value is calculated. In the same Section we derive expressions for the density matrix in the momentum and position representations and discuss its properties. Finally, we present our conclusions and discuss our results in Section \ref{conc}.

We use below the natural Planck system of units setting  Planck and gravitational constants as well as the speed of light to unity.

\section{Basic definitions and equations}
\label{basic}

In what follows we are going to consider the quantum dynamics of a FRW universe with positive spatial curvature, having a cosmological term $\Lambda $, and $n$ massless scalar fields $\varphi_i$, where $i=1,..n$. In this case the WdW 
equation takes the form (e.g. \cite{hm}, \cite{ab})
\begin{equation}
\left ( {\partial^2 \over \partial t^2}-\Delta_n+{\Lambda \over 3}\exp{(6t)}-\exp{(4t)}\right ) \hat \Psi=0,
\label{e1}
\end{equation}
where $t=\ln a$ and $a$ is the scale factor \footnote{Note that often the logarithm of scale factor is denoted as $\alpha$, see e.g. \cite{hm}, \cite{ab}. However, we prefer to use $t$ for this variable to stress its time-like character.}, $\Delta_n=\sum^{n}_{i=1} {\partial^2 \over \partial \varphi_i^2}$ and $\hat \Psi $ is the wavefunction. In agreement with the third quantization procedure we treat the wavefunction as an operator obeying the standard commutation relations:
\begin{equation}
[\hat \Psi(t,x), {\partial \over \partial t}{\hat \Psi}^{\dagger}(t,x^{'})]=i\delta^n(x-x^{'}),
\label{e2}
\end{equation}
where $[..]$ is the field commutator, $x$ represents $n$-dimensional vector with components $\varphi_i$, $\delta^n(x)$ is $n$-dimensional Dirac delta function and the dagger stands hereafter for Hermitian adjoint.

Equation (\ref{e1}) is formally a Klein-Gordon equation for a free quantum field having a time dependant potential
\begin{equation}
V(t)={\Lambda \over 3}\exp{(6t)}-\exp{(4t)},
\label{e3}
\end{equation}
and the associated Lagrangian and Hamiltonian can be written in the form
\begin{equation}
\hat L=\int d^nx \left({\partial \over \partial t}\hat \Psi {\partial \over \partial t}{\hat \Psi}^{\dagger}-
\hat \Psi_{,i}{\hat \Psi_{,i}}^{\dagger}-V\hat \Psi {\hat \Psi}^{\dagger} \right ), \quad \hat H= \int d^nx \left (\hat P {\hat P}^{\dagger}+
\hat \Psi_{,i}{\hat \Psi_{,i}}^{\dagger}+V\hat \Psi {\hat \Psi}^{\dagger} \right ),
\label{e4}
\end{equation}
where commas stand for partial differentiation over $\varphi_i$,
summation over repeating Latin indices is implied from now on, and $\hat P$, ${\hat P}^{\dagger}$ are the canonical momenta. From the Hamilton equations we easily obtain
\begin{equation}
\hat P = {\partial \over \partial t}{\hat \Psi}^{\dagger}, \quad   {\hat P}^{\dagger}={\partial \over \partial t}\hat \Psi.
\label{e5}
\end{equation}

Note that the variable $t$ plays the role of a time-like variable in equation (\ref{e4}), and, therefore, it will be referred to as 'time' below. In order to avoid confusion, let us point out, however, that the proper time is clearly absent in the WdW equation due to its invariance with respect to a choice of the lapse function.

Solutions to (\ref{e1}) can be represented in the standard form
\begin{equation}
\hat \Psi=\int d^nk \left (U_{\omega}e^{ik\cdot x}\hat a_k +U^\ast_{\omega}e^{-ik\cdot x}{\hat b}^{\dagger}_{k} \right ),
\label{e6}
\end{equation}
where $k$ is an $n$ dimensional vector, $\cdot$ stands for the scalar product, $\ast$ is
the complex conjugate and $\omega=\sqrt{k\cdot k}$. The mode functions $U_{\omega}$ satisfy the equation
\begin{equation}
\ddot U_{\omega}+(\omega^2+V)U_{\omega}=0,
\label{e7}
\end{equation}
where dots stand for differentiation over the time-like variable $t$.
Note that we clearly have $U_{\omega}=U_{-\omega}$.

Assuming that the mode functions are normalised according
to the condition
\begin{equation}
U_{\omega}{\dot U_{\omega}}^{\ast}-\dot U_{\omega} U_{\omega}^{\ast}={i\over (2\pi )^n},
\label{e8}
\end{equation}
the operators $\hat a_k$ and $\hat b_k$ obey the standard commutation relations
for the creation and annihilation operators
\begin{equation}
[\hat a_k, {\hat a_{k^{'}}}^{\dagger}]=\delta^n(k-k^{'}), \quad [\hat b_k, {\hat b_{k^{'}}}^{\dagger}]=\delta^n(k-k^{'})
\label{e9}
\end{equation}
with other commutators being equal to zero.

It is worth noting that near the singularity when
$t\rightarrow -\infty$ the potential $V$ tends to zero and equation (\ref{e1}) formally describes
a massless free scalar field in an effective Minkowski spacetime. Thus, when the field mode $U_{\omega}$ is an eigenfunction
of the timelike Killing vector in this spacetime, and, accordingly, $U_{\omega}\propto e^{-i\omega t}$,
the vacuum state $\ket 0$ defined in such a way
that $\hat a_k \ket 0 =0$ and  $\hat b_k \ket 0 =0$ for all $k$ is the standard vacuum state for a massless scalar field in this asymptotic limit. We use this state as the field state in our analysis below.

For our purposes it is sometimes convenient to use another representation for  the field $\hat \Psi$ through a Fourier transform
\begin{equation}
\hat \Psi =\int d^nk e^{ik\cdot x}\hat \Psi_{\omega}, \quad \hat \Psi_{\omega}={1\over \sqrt 2}\left(U_{\omega}(\hat c_{1,k}+i\hat c_{2,k})+
U_{\omega}^\ast({\hat c_{1,k}}^{\dagger}+i{\hat c_{2,k}}^{\dagger})\right),
\label{e10}
\end{equation}
where we introduce new creation and annihilation operators
\begin{equation}
\hat c_{1,k}={1\over \sqrt 2}(\hat a_{k}+\hat b_{-k}), \quad \hat c_{2,k}={i\over \sqrt 2}(\hat b_{-k}-\hat a_{k}).
\label{e11}
\end{equation}
It is easy to see that these operators obey the standard commutations. Also, it is evident that the vacuum
state $\ket 0$ defined above is also a vacuum with respect to these operators.

The Hamiltonian (\ref{e4}) can be expressed in terms of $\hat \Psi_{\omega}$ as
\begin{equation}
\hat H=(2\pi)^n\int d^nk\left ( \hat P_{\omega} {\hat P_{\omega}}^{\dagger}+
({\omega}^2+V)\hat \Psi_{\omega} {\hat \Psi_{\omega}}^{\dagger}\right ), \quad \hat P_{\omega}={\partial \over \partial t}{\hat \Psi_{\omega}}^{\dagger}.
\label{e12}
\end{equation}
It can be brought to a standard form by separating $\hat \Psi_{\omega}$ into the real and imaginary parts
as $\hat \Psi_{\omega} ={1\over \sqrt{2(2\pi)^n}}(\hat q_{1,\omega} + i\hat q_{2,\omega})$, where $\hat q_{1,\omega} $ and
$\hat q_{2,\omega}$ are Hermitian and commute with each other. We have
\begin{equation}
\hat H=\sum_{\alpha=1,2} {1\over 2}\int d^nk\left( {\hat p_{\alpha,\omega}}^2+
({\omega}^2+V){\hat q_{\alpha, \omega}}^2\right), \quad \hat p_{\alpha, \omega}={\partial \over \partial t}
\hat q_{\alpha, \omega}.
\label{e13}
\end{equation}
The expression (\ref{e13}) tells that the problem can be formulated in terms of an infinite set of oscillators with a time dependent frequency. Based on the analogy with the oscillators we introduce other time-dependent ``creation and annihilation operators'', $\hat d_k(t)$ and ${\hat d_k}^{\dagger}(t)$,
according to the rule
\begin{equation}
\hat d_k(t)={1\over \sqrt 2}\left(\sqrt{\Omega (t_0)}\hat q_k + {i\over \sqrt{ \Omega (t_0)}}\hat p_k\right),
\label{e14}
\end{equation}
where
\begin{equation}
\Omega (t_0)=\sqrt{\omega^2+V(t_0)},
\label{en14}
\end{equation}
the potential $V$ is assumed to be taken at a particular fixed moment of time, $t_0$, and, therefore, $\Omega (t_0)$ does not
depend on time $t$. Also, for simplicity we assume hereafter $t_0$ is such that $\omega^2+V(t_0) > 0$, and,
accordingly, $\Omega (t_0)$ is real.
By construction the operators (\ref{e14}) obey the standard commuting relations. Thus, they can be related to the operators
$\hat c_k$ by a Bogolyubov transformation
\begin{equation}
\hat d_k(t)=\alpha_{\omega} \hat c_k + \beta_{\omega} {\hat c_k}^{\dagger}, \quad |\alpha_{\omega}|^2-|\beta_{\omega}|^2=1.
\label{e15}
\end{equation}
The explicit form of the Bogolyubov coefficients follows from expressions (\ref{e10}-\ref{e14}):
\begin{equation}
\alpha_{\omega}=\sqrt{{(2\pi)^n\over 2}}\left(\sqrt{\Omega (t_0)} U_{\omega}+{i\over \sqrt{\Omega (t_0)} }\dot U_{\omega}\right),
\quad \beta_{\omega}=\sqrt{{(2\pi)^n\over 2}}\left(\sqrt{\Omega (t_0)} U^\ast_{\omega}+{i\over \sqrt{\Omega (t_0)} }\dot U^\ast_{\omega}\right).
\label{e16}
\end{equation}

The normalisation condition (\ref{e8}) tells that the Bogolyubov coefficients do obey the second equality in (\ref{e15}).

The vacuum state defined by the condition $d_k(t_0)\ket 0_{ad} =0$ defines so-called ``adiabatic'' vacuum state $\ket 0_{ad}$ (see e.g \cite{bd}). Note that, by definition, this state does not depend on time.

Of special importance are two particular values of $t_0$, $t_0 \rightarrow \pm \infty$.

In the former case the adiabatic vacuum state provides a natural $out$-vacuum state for our
problem. Thus, when $\beta_k (t_0 \rightarrow \infty) \ne 0$ the state $\ket 0$ contains excitations above the $out$-vacuum usually interpreted as ``creation of particles'' (or universes, in our
case) from 'nothing'.

The latter case corresponds to setting $V(t_0)$ in (\ref{en14}) to zero. Clearly, in this situation relations (\ref{e15}) and (\ref{e16}) simply determine the evolution of $\hat c_k$ and  ${\hat c_k}^{\dagger}$ in the Heisenberg representation. On the other hand, in the Schrodinger representation these relations can be used to determine the evolution of the wave functional, see e.g. \cite{pol} and Section \ref{wigner} below.

\section{An asymptotic analytic solution of the WdW equation}
\label{wdw}

In general, solutions to equation (\ref{e7}) can be obtained analytically only when either the first or the second term in the expression for potential $V$ in (\ref{e3}) is discarded. This corresponds to neglecting the influence of the $\Lambda $ term or spatial curvature, respectively. When both terms in equation (\ref{e3}) are retained it can either be solved numerically or an approximate solution can be looked for using appropriate asymptotic methods. Here we consider the second possibility in detail in the limit $\Lambda \ll 1$, which may be appropriate for inflationary models, where the value of the $\Lambda $ term is generally much smaller than its natural Planck value $\Lambda = 1$. Numerical solutions will be used to validate our analytic approach.

Additionally, for simplicity we are going to consider sufficiently small values of the frequency $\omega$, namely, in our analytical work we formally assume hereafter that $\omega \ll \omega_{crit}$, where $\omega_{crit}=\sqrt{-V_{min}(t_{min})}={2\over \sqrt 3}{1\over \Lambda}$,
and $V_{min}(t_{min})=-{4\over 3\Lambda^2}$, $t_{min}={1\over 2}\ln {{2\over \Lambda}}$ are the minimal (negative)
value of the potential $V$ and the corresponding value of the variable $t$, respectively. For such
values of $\omega $ we can use the WKBJ approximation, assuming that a solution to (\ref{e7}) is approximately
proportional to
\begin{equation}
{1\over \sqrt{\dot S}}e^{i S(t)},\quad  S=\pm \int dt \sqrt {\omega^2+V}, \quad |S| \gg 1
\label{e17}
\end{equation}
both in the classically forbidden region corresponding to an intermediate range of $t$, where $\omega \ll -V$ and in the classically allowed regions corresponding to small values of $t$, where $V \rightarrow 0$, and large values of $t$, where $V \gg \omega$.
In the forbidden region the phase $S$ is purely imaginary, and therefore a general solution to (\ref{e7}) has
the form of a sum of growing and decaying exponents multiplied by the pre-exponential factor, while
in the allowed region $S$ is real and the solution to  (\ref{e7}) has oscillatory behaviour. It is very
important to note that the WKBJ approximation breaks down when the time $t$ is close to $t_\ast= {1\over 2}\ln {{3\over \Lambda}}$ such that $V(t_\ast)=0$. However, in the vicinity of this moment of time one can simplify
(\ref{e7}), find an appropriate exact solution of the simplified equation and match it to the WKBJ solution.

\begin{figure}[htbp]
	\begin{center}
		\includegraphics[width=0.7\textwidth]{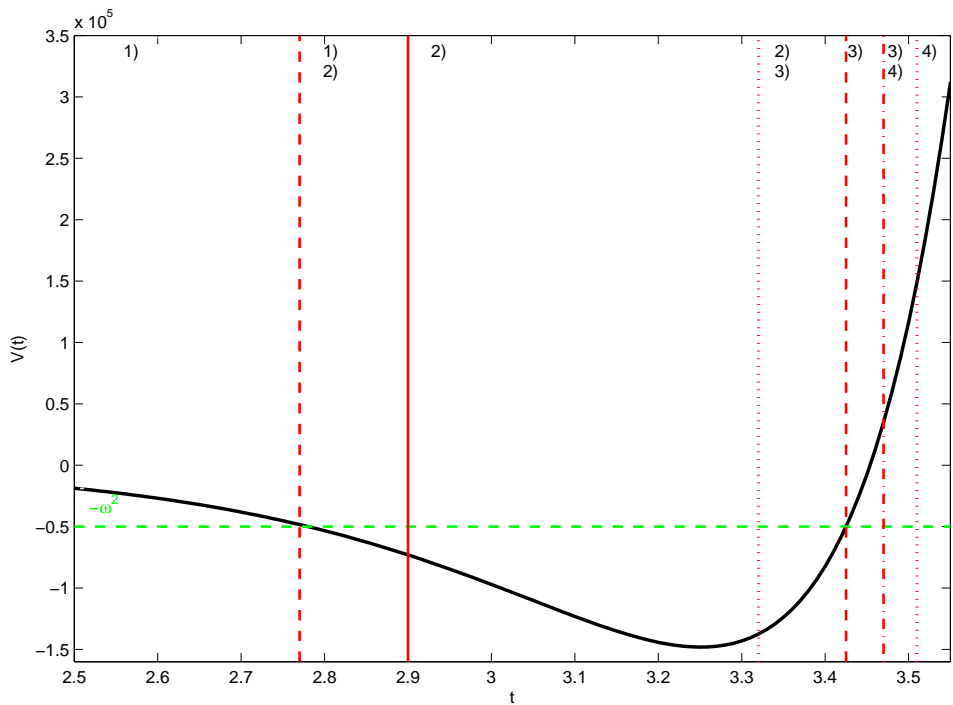} 
		\caption{We show the potential $V(t)$ together with four overlapping regions, where various approximations discussed in the text are
possible. The potential is calculated for $\Lambda=3*10^{-3}$, square of the mode frequency is taken to be $\omega^2=0.5$. Since the precise
positions of boundaries of these regions are ambiguous they are shown schematically. Vertical solid, dashed, dotted, dot-dashed lines
show the positions of boundaries of regions 1), 2), 3) and 4), respectively.}
		\label{fig1}
	\end{center}
\end{figure}

Thus, the whole time interval $-\infty < t < +\infty $ can be subdivided into four overlapping regions: 1) the region of sufficiently small $t$, where the curvature term dominate over the term proportional to $\Lambda $ in the expression for the potential, 2) the classically forbidden region, 3) the region close to the moment $t=t_\ast$ and 4) the classically allowed region. These regions together with the potential $V(t)$ are schematically shown in Fig. \ref{fig1}. Matching solutions in all these regions and using the normalisation condition (\ref{e8}) we can find a solution for $U_{\omega}$ approximately valid when $\omega \ll \omega_{crit}$. As we shall see below, a comparison with numerical results shows that this solution is qualitatively valid even when $\omega \le \omega_{crit}$. On the other hand, an approximate solution in the opposite limit $\omega \gg \omega_{crit}$ can be obtained by setting the term determined by the curvature in (\ref{e3}) to zero. Its form is well known, see e.g. \cite{kif} and unimportant for us.

Let us consider the behaviour of $U_{\omega}$ by turns, starting from region 1).

In this region the explicit form of the solution can be written as
\begin{equation}
U_{\omega} \approx C_{\omega} I_{-i{\omega \over 2}}\left ({e^{2t}\over 2}\right ), \quad C_{\omega}=2^{-i\omega}{\Gamma \left(1-{i\omega \over 2}\right)\over \sqrt{2(2\pi)^{n}\omega}},
\label{e18}
\end{equation}
where $I_{\nu}(x)$ is the modified Bessel function of the first kind, the coefficient $C_{\omega}$ is determined by the normalisation condition (\ref{e8}) and $\Gamma (x)$ is the gamma function, see also e.g. \cite{kim}. 
It is easy to see that when $e^{2t} \ll 1$ from equation (\ref{e18}) it follows that $U_{\omega} \approx   {1\over \sqrt{2(2\pi)^{n}\omega}}e^{-i\omega t}$, which is just the standard positive frequency solution in $(n+1)$-dimensional Minkowski spacetime normalised according to (\ref{e8}), see e.g. \cite{ab}.

Below we shall need an asymptotic form of the modified Bessel function of purely
imaginary index at large values of its argument. As discussed in e.g. \cite{dun} to this purpose
it is convenient to consider the modified Bessel function of second kind $K_{\nu}(x)$ and an additional
function $L_{\nu}(x)$ defined according to the rule
\begin{equation}
K_{\nu}(x)={\pi \over 2\sin (\pi \nu)}\left(I_{-\nu}(x)-I_{\nu}(x)\right ), \quad L_{\nu}(x)={i \pi \over 2\sin (\pi \nu)}\left(I_{-\nu}(x)+I_{\nu}(x)\right).
\label{en18}
\end{equation}
Note that both functions are real when their argument is real and the index is purely imaginary. As shown in \cite{dun} when $x \rightarrow \infty $, $L_{i\mu}(x) \approx {1\over \sinh(\pi \mu)}\sqrt {\pi \over 2x}e^x$, while $K_{i\mu}(x)$ has the well known asymptotic form: $K_{i\mu}(x)\approx \sqrt {\pi \over 2x}e^{-x}$ in the same limit. Using these expressions and considering sufficiently large values of $t$ we obtain from (\ref{e18}) and (\ref{en18})
\begin{equation}
U_{\omega} \approx {C_{\omega}\over \sqrt{\pi}}e^{-t}\left (\exp \left({e^{2t}\over 2}\right) + i\sinh\left({\pi \omega \over 2}\right)
\exp \left(-{e^{2t}\over 2}\right) \right ).
\label{enn18}
\end{equation}
Note that although the last term in the brackets is exponentially small at large $t$ it is needed to be retained for the asymptotic solution (\ref{enn18}) to satisfy the normalisation condition (\ref{e8}).

The solution (\ref{enn18}) can be matched to a WKBJ solution of the form (\ref{e17}) in a time interval within region 2), where on one hand the time $t$ is small enough such that $t <  t_{min}$ , and on the other hand, it is sufficiently large for the condition $\omega \ll \sqrt{-V}$ to be fulfilled.

Before doing so let us discuss the WKBJ phase, $S$, in (\ref{e17}). In general, it can be represented in the form
\begin{equation}
 S=\pm i \int {dx \over 2x} \sqrt {P(x)}, \quad P(x)=4x^2-{8\over 3}\Lambda x^3-\omega^2,
\label{e19}
\end{equation}
where we introduce a new independent variable $x={e^{2t}\over 2}$. Since the polynomial $P(x)$ is of the third order
explicit integration over $x$ in (\ref{e19}) is rather cumbersome. It becomes trivial, however, when we
set $\omega=0$. In this case  $P(x)\propto x^2$, and the integration gives
$S=\mp {i\over \Lambda}(1-{2\Lambda x\over 3})^{3/2}$. In order to avoid unnecessary complications let us
take into account that in the limit $\omega \ll \omega_{crit}$ the term $\omega^2$ in (\ref{e19}) may play
a role only when $t \sim t_*$. To take this fact into account we can consider a modified polynomial,
$P_{mod}$, in (\ref{e19}) fixed by the following conditions: it is of third order,
$P_{mod}\propto x^2$, $P_{mod}=P$ when $\omega=0$ and $P_{mod}(t_\ast)=0$.
By doing so we obtain $P_{mod}=x^2(4-{4\over 9}\Lambda ^2 \omega^2-{8\over 3}\Lambda x)$ and $S=\mp  {i\over \Lambda}(1-{\Lambda^2 \omega^2\over 9}-{2\Lambda x\over 3})^{3/2}$.

Taking this into consideration it is easy to see that the matching procedure in region 2) gives in the leading order
\begin{equation}
 U_{\omega}\approx {C_{\omega}\over \sqrt \pi}{e^{-t}\over \phi_-^{1/4}}\left (e^{{1\over \Lambda}(1 -\phi_-^{3/2})}+
i\sinh\left({\pi \omega\over 2}\right)e^{-{1\over \Lambda}(1 -\phi_-^{3/2})} \right ),
\label{e20}
\end{equation}
where
\begin{equation}
\phi_-=1-{\Lambda^2 \omega^2\over 9}-{\Lambda \over 3}e^{2t}.
\label{en20}
\end{equation}

Now let us consider region 3), where $t \sim t_\ast$. For that we assume that relative
difference $(t_\ast-{\omega^2 \Lambda^2\over 18}-t)/t_\ast$ is small and simplify equation (\ref{e7}) retaining only leading order terms with respect to the relative difference. It is convenient to introduce a natural variable $z={18^{1/3}\over
\Lambda^{2/3}}(t_\ast-{\omega^2 \Lambda^2\over 18}-t)$ to see that after the simplification equation (\ref{e7})
is reduced to the Airy equation
\begin{equation}
{d^2\over d z^2}U_k - zU_k=0.
\label{e21}
\end{equation}
Solutions to (\ref{e21}) are matched to (\ref{e20}) in the region, where both conditions
$(t_\ast-{\omega^2 \Lambda^2\over 18}-t)/t_\ast \ll 1$
and $z \gg 1$ are fulfilled. This gives the solution in the form
\begin{equation}
U_{\omega}=C_{\omega}2^{-1/6}\left ({ \Lambda \over 3}\right )^{1/3}\left (2e^{{1\over \Lambda}} Ai(z)+i
\sinh\left({\pi \omega\over 2}\right)e^{-{1\over \Lambda}}Bi(z)\right ),
\label{e22}
\end{equation}
where $Ai(z)$ and $Bi(z)$ are Airy functions of the first and the second kind, respectively.

The solution can be analytically continued to the region $z < 0$. Using
a standard result and assuming that $z_1\equiv -z \gg 1$ we obtain
\begin{equation}
U_{\omega}=C_{\omega}{2^{-1/6}\over \sqrt{\pi}z_1^{1/4}}\left ({ \Lambda \over 3}\right )^{1/3}
\left (2e^{{1\over \Lambda}}\sin\left({2\over 3}z_1^{3/2}+{\pi\over 4}\right)+ i
\sinh\left({\pi \omega\over 2}\right)e^{-{1\over \Lambda}} \cos\left({2\over 3}z_1^{3/2}+{\pi\over 4}\right)\right ).
\label{e23}
\end{equation}

Finally, we can match solution (\ref{e23}) to the solution of the form (\ref{e17}) in the
classically allowed region 4) using the same technique as the one leading to expressions (\ref{e22}) and (\ref{e23}) and taking into account
that in this region $S={1\over \Lambda}({\Lambda \over 3}e^{2t}+{\Lambda^2 \omega^2\over 9}-1)^{3/2}$.
We have
\begin{equation}
U_{\omega}={C_{\omega}\over \sqrt \pi} {e^{-t}\over \phi^{1/4}_+}\left (2e^{{1\over \Lambda}}\sin\left({1\over \Lambda}\phi_+^{3/2}+{\pi\over 4}\right)+ i
\sinh\left({\pi \omega\over 2}\right)e^{-{1\over \Lambda}} \cos\left({1\over \Lambda}\phi_+^{3/2}+{\pi\over 4}\right)\right ),
\label{e24}
\end{equation}
where
\begin{equation}
\phi_+\equiv -\phi_-={\Lambda \over 3}e^{2t}+{\Lambda^2 \omega^2\over 9}-1.
\label{e25}
\end{equation}

Equations (\ref{e18}), (\ref{e20}), (\ref{e22}) and (\ref{e24}) give the expression of $U_{\omega}$ in regions 1)-4), respectively. Note that the term $\propto \omega^2$ in (\ref{e25}) must be discarded in the limit $t \rightarrow \infty$, since it gives an artificially large contribution due to the use of $P_{mod}(x)$ instead of $P(x)$ in our analysis above.

\section{Transition to quasi-classicality}
\label{trans}

Equation (\ref{e7}) describes an oscillator with a time-dependent frequency. Its form is analogous to the form of the equation used, for example, to study the evolution of a test massless scalar quantum field in an expanding Universe, see e.g. \cite{star}, \cite{lindeb} \footnote{Note that equations describing the evolution of scalar and tensor cosmological perturbations also have the same structure, see e.g. \cite{luk} and \cite{gr1}.}, although potential $V_c(\eta)$ used in the cosmological setting has another time dependence\footnote{Note that in the cosmological context the potential is often defined with the opposite sign.}, $\eta$ is the so-called 'conformal time'. In a class of cosmological models, notably in the important case of inflationary models $V_c(\eta)$ tends to zero at small values of $\eta$, while at  sufficiently large $\eta $ its absolute value can be much larger than the square of the wavenumber, $k_c^2$, characterising a particular field mode. In this case the $in$-vacuum state specifying initial conditions for the field evolution is typically defined in a way analogous to what is used in this paper. Modes evolving in the regime $k_c < \sqrt{|V_c(\eta)|}$ during a considerable period of time experience a copious growth of occupation numbers with respect to a natural $out$-vacuum state. When inflationary models are considered, such a regime corresponds to physical wavelengths of the modes being larger than the cosmological horizon scale, and, accordingly, $k_c \ll 1/\eta$. In the framework of these models it has been shown (see e.g. \cite{star}, \cite{pol}) that a large scale part of the field consisting of the modes with sufficiently small wavenumbers, which have been evolved outside the cosmological horizon for a long time behave as a classical random field with zero expectation value. We suppose similar behaviour in our case for the part of the wavefunction consisting of modes with $\omega < \omega_{crit}$, at large values of our time variable $t$. To probe the classical behaviour we are going to calculate explicit values of the Bogolyubov coefficients given by expressions (\ref{e16}) and, following \cite{pol}, the time dependence of the Wigner function corresponding to our quantum state, which has a special form for a system evolving in a quasi-classical regime.

\begin{figure}[htbp]
	\begin{center}
		\includegraphics[width=0.6\textwidth]{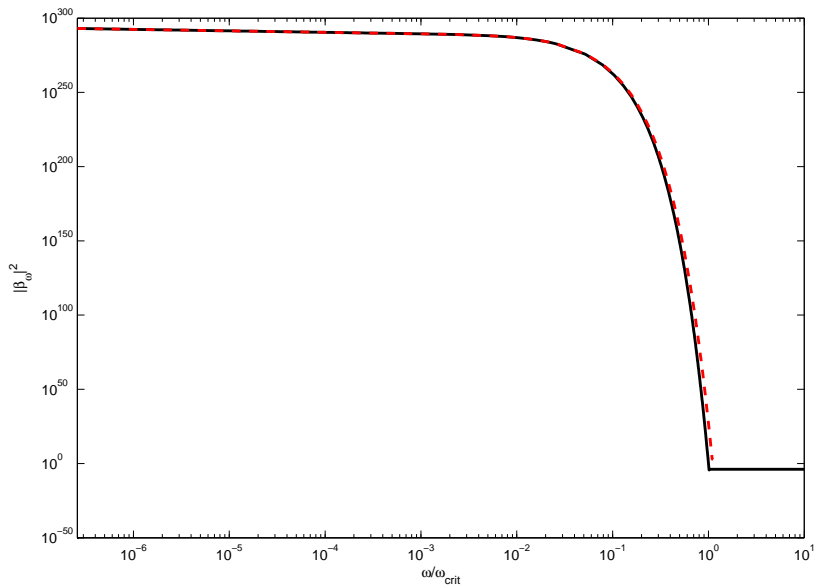} 
		\caption{The square of absolute value of the Bogolyubov coefficient $\beta_{\omega}$ is shown as a function of
ratio $\omega/\omega_{crit}$, for $\Lambda = 3*10^{-3}$. Solid and dashed curves represent  numerical result and expression (\ref{e27}), respectively. }
		\label{fig2}
	\end{center}
\end{figure}

\subsection{Calculation of the amount of produced universes }
\label{calc}

The amount of universes produced at late 'times' $t \rightarrow \infty$ per unit of volume in $k$ space
\footnote{In the classically allowed region the phase $S$ plays the role of classical action
characterising a particular classical trajectory of our model.
Using this fact, one can show that every component of vector $k$ is proportional to
the derivative of corresponding field $\phi_i$  over the physical time.}
is given by the square of an absolute value of Bogolyubov coefficient $\beta$,
\begin{equation}
|\beta_{\omega}|^2={(2\pi)^n\over 2}\left(\Omega(t_0) U_{\omega}U_{\omega}^{*}+\Omega^{-1}(t_0)\dot U_{\omega}{\dot U_{\omega}}^{*}\right)-1/2,
\label{e26}
\end{equation}
where we use (\ref{e8}) and (\ref{e16}). Note that the physical meaning of multiple production of universes was discussed in e.g. \cite{hm}. As discussed in this paper  $|\beta|^2$ may be considered as being proportional to a probability to find a universe with given parameters, in our case characterised by different values of $k$. In quantum cosmology the picture of multiple universes itself has a direct physical meaning either when their wavefunctions can interfere (e.g. \cite{hm}) or through non-linear interaction terms added to the quantized WdW equation, see e.g. \cite{mg}.

In order to obtain $|\beta|^2$ explicitly we assume that $t=t_0\rightarrow \infty $, thus calculating occupation numbers with respect to our $out$-vacuum state. In this limit
we have $\Omega (t_0) \approx \sqrt{{\Lambda \over 3}}e^{3t}$ and $\phi_+\approx  {\Lambda \over 3}e^{2t}$. Substituting
these expressions into (\ref{e24}) and the resulting one into (\ref{e26}) we obtain
\begin{equation}
|\beta_{\omega}|^2={1\over 8 \sinh\left({\pi \omega\over 2}\right)}\left (2e^{{1\over \Lambda}}-\sinh\left({\pi \omega\over 2}\right)e^{-{1\over \Lambda}}\right )^2
\label{e27}
\end{equation}
\footnote{In the course of the calculation we use the well known relation $\Gamma (1+ix)\Gamma (1-ix)={\pi x\over
\sinh (\pi x)}$.}.
Equation (\ref{e27}) tells that $|\beta_{\omega}|^2\propto e^{{2\over \Lambda}}$  when $\omega \ll \omega_{crit}$. This result was obtained in \cite{rub} by qualitative means. On the other hand, as it is evident from  (\ref{e27}) $|\beta_{\omega}|^2$ strongly decreases with increase of $\omega$. It is formally equal to zero at $\omega \approx
{4\over \pi \Lambda}$. However, this value is slightly larger than $\omega_{crit}$ and we
derived  (\ref{e27}) assuming that $\omega \ll \omega_{crit}$. Therefore, the latter effect may be an artifact of our approximations.  Fig. \ref{fig2}  shows $|\beta_{\omega}|^2$ given by equation (\ref{e27}) as well as the result of calculation of $|\beta_{\omega}|^2$ by numerical means. One can see that there is a very good agreement between analytic and numerical results at small values of $\omega$. When $\omega \sim \omega_{crit}$ numerical
and analytic curves differ by several orders of magnitude, but have qualitatively similar behaviour. In both cases there is a drastic decrease of $|\beta_{\omega}|^2$ towards $\omega_{crit}$. When $\omega > \omega_{crit}$ the numerical curve shows that $|\beta_{\omega}|^2$ is small.

\subsection{Wigner function}
\label{wigner}

As was discussed in e.g. \cite{gr2}, \cite{pol} in  a quasi-classical regime the Wigner function has a special form of a distribution over a generalised coordinate multiplied by a sharply peaked distribution centred at a linear combination of generalised momentum and coordinate of a classical trajectory. Let us calculate it for our model. In this Section, we assume, for simplicity, that our quantum field $\Psi^{\dagger}$ is quantized in a box, since this assumption doesn't influence our conclusions. This has an advantage that we shall deal with ordinary functions rather than functionals when treating
wavefunctions corresponding to the quantum state of our system\footnote{Let us stress the difference between a wavefunction corresponding to a quantum state of our system and the wavefunction of the Universe. The latter plays the role of a generalised coordinate in the formalism of third quantization.}.  In this case the Kronecker deltas are implied in (\ref{e2}) and (\ref{e8}) instead of delta functions, and summation instead of integration in the expressions containing integrals over $k$ is to be used. The normalisation condition also changes but we are nonetheless going to use equation (\ref{e8}) since  it doesn't affect our results.

After this assumption is made, the expression  (\ref{e13}) tells that our model is reduced to a discrete set of oscillators with
time dependent frequency and different values of $k$ and $\alpha$. Let us consider one of them with a particular value of
$\omega $. The Wigner function corresponding to this oscillator has the standard form
\begin{equation}
W(p,q)={1\over \pi}\int^{\infty}_{-\infty} dy \Psi_{S}^{\ast}(q+y)\Psi_{S}(q-y)e^{2ipy},
\label{e28}
\end{equation}
where c-numbers $p$ and $q$ are conjugated momentum and coordinate, $\Psi_{S}(q)$ is the wave function in the Schrodinger representation,
and we omit indices $k$, $\alpha $ and $\omega$ in this Section.

To find the evolution of $\Psi_{S}(q)$ let us note that in the Schrodinger representation the operators $\hat p$ and $\hat q$ entering (\ref{e13}) have the standard form $\hat p =-i{\partial \over \partial q} $ and $\hat q = q$. On the other hand  $\Psi_{S}(q)$ corresponds to the vacuum state, and, therefore, $\hat c \Psi_{S}(q)=0$. Setting  $\Omega (t_0)=\omega$ in equations (\ref{e14})-(\ref{e15}) we express $\hat c$ in terms of $\hat p$ and $\hat q$ to get
\begin{equation}
\left({d\over dq}+ Dq\right)\Psi_S(q)=0,
\label{e29}
\end{equation}
where
\begin{equation}
D=\omega \left({\alpha^\ast-\beta \over \alpha^\ast + \beta}\right).
\label{e30}
\end{equation}
Note that the same equation was obtained in \cite{pol} in the quantum field context.
Taking into account that when $t\rightarrow -\infty $ $\Psi_{S}(q)$ tends to that of the vacuum state of an oscillator with frequency $\omega$
we have
\begin{equation}
\Psi_S(q)=\left({K\over 2\pi}\right)^{1/4}\exp \left(-{Dq^2\over 2}\right), \quad K=D+D^\ast,
\label{e31}
\end{equation}
and, substituting (\ref{e31}) into (\ref{e28}) we obtain
\begin{equation}
W(p,q)={1\over \pi}\exp \bigg[-\left({Kq^2\over 2}+{(2p+Rq)^2\over 2K}\right)\bigg], \quad R=i(D^*-D).
\label{e32}
\end{equation}
The Wigner function has an important symmetry with respect to interchange of the canonical variables. Namely, it can
be written in another form
\begin{equation}
W(p,q)={1\over \pi}\exp \bigg[-\left({K_np^2\over 2}+{(2q-R_np)^2\over 2K_n}\right)\bigg], \quad K_n={1\over D^*}+{1\over D},
\quad R_n=i\left({1\over D^*}-{1\over D}\right).
\label{e33}
\end{equation}
This property follows from the observation that when the momentum rather than coordinate
representation for the operators $\hat p$ and $\hat q$ is
used we have a similar equation for wavefunction differing from (\ref{e29}) by a new coefficient $D_n=1/D$. Also, in the momentum representation there should be the opposite sign of the argument in the exponent in (\ref{e28}).
Note that since the Wigner function is positive definite in our case we treat it as determining
a probability distribution in phase space.

When either $K\rightarrow 0 $ or $K_n\rightarrow 0$  distributions (\ref{e32}), (\ref{e33}) are sharply
peaked around $2p+Rq=0$ and $2q-R_np$, respectively. Let us show that these relations hold ``on average ``
for a classical solution of our equations of motion in the limit $t \gg 1$ and
calculate coefficients $K$, $K_n$, $R$ and $R_n$ explicitly, in the same limit.
To do so we simplify the expressions for $\alpha $ and $\beta $ in equation (\ref{e16}) using
the same approximations as in Section \ref{calc}, but now setting $\Omega(t_0)=\omega$ there. We obtain
\begin{equation}
\alpha=(2\pi)^{{n-1\over 2}}C_{\omega} \left((\omega^{1/2}\Omega^{-1/2}(t)a_1+\omega^{-1/2}\Omega^{1/2}(t)a_2)\sin \phi
+i(\omega^{1/2}\Omega^{-1/2}(t)a_2+\omega^{-1/2}\Omega^{1/2}(t)a_1)\cos \phi \right ),
\label{e34}
\end{equation}
and
\begin{equation}
\beta=(2\pi)^{{n-1\over 2}}C^{\ast}_{\omega}\left ((\omega^{1/2}\Omega^{-1/2}(t)a_1-\omega^{-1/2}\Omega^{1/2}(t)a_2)\sin \phi
+i(\omega^{1/2}\Omega^{-1/2}(t)a_1-\omega^{-1/2}\Omega^{1/2}(t)a_2)\cos \phi \right ),
\label{e35}
\end{equation}
where, by definition, $\Omega (t)=\sqrt{{\Lambda \over 3}}e^{3t}$ is the asymptotic value of (\ref{en14}) with $t_0$
being substituted by $t$, in the limit $t \rightarrow
\infty$, $a_1=2e^{{1\over \Lambda}}$, $a_2=\sinh ({\pi \omega \over 2})e^{-{1\over \Lambda }}$, $\phi = {\Lambda^{1/2}\over 3^{3/2}}e^{3t}(1-{3\over \Lambda}e^{-2t})^{3/2}+{\pi \over 4}$
is the phase of sine and cosine entering (\ref{e24}). Note that $\dot \phi \approx \Omega (t)$ and we use equation (\ref{e25}) for $\phi_+$ setting
$\omega=0$ there due to the reason explained above, but taking into account the curvature term in the arguments of sine and cosine.

Substituting (\ref{e34}) and (\ref{e35})
in (\ref{e30})-(\ref{e33}) we have
\begin{equation}
K=2\Omega {a_1a_2\over a_1^2\sin^2 \phi +a_2^2\cos^2 \phi}\approx 2\Omega {a_2\over a_1}{1\over \sin^2 \phi}, \quad
R=\Omega {(a_2^2-a_1^2)\sin 2\phi \over  a_1^2\sin^2 \phi +a_2^2\cos^2 \phi}\approx -2\Omega \cot \phi ,
\label{e36}
\end{equation}
and
\begin{equation}
K_n=2\Omega^{-1} {a_1a_2\over a_2^2\sin^2 \phi +a_1^2\cos^2 \phi}\approx 2\Omega^{-1} {a_2\over a_1}{1\over \cos^2 \phi}, \quad
R_n=\Omega^{-1} {(a_1^2-a_2^2)\sin 2\phi \over  a_2^2\sin^2 \phi +a_1^2\cos^2 \phi}\approx 2\Omega^{-1} \tan \phi ,
\label{e37}
\end{equation}
where $\Omega \equiv \Omega (t)$ from now on,
we take into account that when $\omega \ll \omega_{crit}$ we have $a_2 \ll a_1$ to get the approximate expressions.
Note that these are valid only when the phase $\phi$ is not
very close to $\phi_j=\pi j$ and $\phi_j=\pi/2+\pi j$ ($j$ is an integer), respectively, for equations (\ref{e36}) and
(\ref{e37}). Hereafter, we assume that the value of time variable
$t$ is such that this condition is satisfied. The opposite case should be treated separately.

When the classical motion is considered $p=\dot q$. Using this fact and the approximate expressions for $R$ and $R_n$
in (\ref{e36}) and (\ref{e37}) we easily see that both combinations $2p+Rq$ and $2q-R_np$ are equal to zero provided that
$q$ is proportional to $\sin \phi$, which is an approximate solution of the classical equations of motion at large times. Thus, in both cases
of small $K$ and $K_n$ distributions (\ref{e32}) and (\ref{e33}) can be treated as describing bundles of classical trajectories
with a random Gaussian distribution of the coordinate and momentum, respectively. It is interesting to point out that
these trajectories have the same phase $\phi$. Thus, in the quasi-classical limit only one of two linearly independent
solutions of the equations of  motion is present. This is analogous to the test
quantum field problem, where the so-called 'growing'
mode is singled out in the same regime. Let us stress again that this picture is not correct when the degenerate set of
phase $\phi_j$ is considered. In the latter case the behaviour of the system is no longer quasi-classical.

It is instructive to introduce a 'nascent delta function' $\delta_{\epsilon}(x)={1\over \sqrt{2\pi \epsilon}}e^{-{{x^2\over 2\epsilon}}}$ having the
property that $\delta_{\epsilon \rightarrow 0}(x)\rightarrow \delta (x)$ and rewrite (\ref{e32}) and (\ref{e33}) in an equivalent form as
\begin{equation}
W(p,q)=\sqrt{{K\over 2\pi}}e^{-{Kq^2\over 2}}\delta_{K}\left(p+{R\over 2}q\right)=\sqrt{{K_n\over 2\pi}}e^{-{K_nq^2\over 2}}\delta_{K_n}\left(q-{R_n\over 2}p\right).
\label{en35}
\end{equation}
Equation (\ref{en35}) tells that when $K\rightarrow 0$ or $K_n\rightarrow 0$ the probability distribution over $q$ or $p$ has
a dispersion much larger than $1$, while the probability distribution over the other coordinate shrinks. This is a characteristics
of a strongly squeezed vacuum state. It is well known in this case the coordinate having a large dispersion can be treated as a classical random quantity with a Gaussian distribution \cite{gr2}.Thus, we have two possible cases of quasi-classical evolution
of one particular mode referred to hereafter as case one and case two. In the former case corresponding to modes with relatively small $\omega $ the coordinate can be treated as a
classical random variable, while in the latter case valid for modes with relatively large $\omega$, which are, however, smaller than $\omega_{crit}$  it is the momentum, which is quasi-classical.

From the condition $\bar {1\over K}\equiv {1\over \pi}\int d\phi {1\over K} =1$ let us find a typical frequency,
$\omega_s$, which separates modes evolving in the two different quasi-classical regimes. When $\omega \ll
\omega_s$ we have case one, while the opposite case two is realised when $\omega_s \ll \omega \ll \omega_{crit}$.
Note that when $\omega \sim \omega_s$ the system behaviour is
essentially quantum. We have
\begin{equation}
\omega_s ={2\over \pi}\left ({2\over \Lambda }-3t-{1\over 2}\ln \left({\Lambda \over 3}\right)\right ),
\label{e38}
\end{equation}
where we assume, for simplicity, that ${\pi \omega_s\over 2} > 1$. Neglecting the logarithmic correction
we see that the case one is present in the system only when
\begin{equation}
t < t_{crit}\equiv {2\over 3\Lambda },
\label{e39}
\end{equation}
and, accordingly, the scale factor $a < a_{crit}=e^{{2\over 3\Lambda }}$.

\section{Quasiclassical stochastic wavefunction of the Universe and its description through a density matrix}
\label{wave}
\subsection{Quasiclassical wavefunction}
\label{wave1}
As we discussed in the previous Section when $1 \ll t \ll t_{crit}$
the large scale part of the $\hat \Psi$ consisting of  modes with $\omega \ll \omega_s$, $\Psi_{qc}$, may be considered as a quasi-classical one. It can be represented using an approach analogous to the so-called coarse graining procedure frequently used for quantum fields evolving in an inflationary Universe \cite{star} as an integral over all modes with frequencies smaller
than $\omega_{L}=\xi_L \omega_s$, where a constant $\xi_L $ is assumed to be small. It is easy to see that $\omega_L$ is given by the same expression as (\ref{e38}), but with the argument of the logarithm being divided by $\xi_L$:  $\sqrt {{\Lambda \over 3}}\rightarrow
\sqrt {{\Lambda \over 3}}\xi_L^{-1}$. We have
\begin{equation}
\Psi_{qc}=\int d^nk \big (U_{\omega}e^{ik\cdot x}a_k +U^\ast_{\omega}e^{-ik\cdot x}{b}_k^{\ast} \big ),
\label{e40}
\end{equation}
where all quantities are assumed to be c-numbers and the mode function is given by (\ref{e24}).
Complex random numbers $a_k$ and $b_k$ have Gaussian distributions
and must be normalised in such a way that resulting statistical averages of different products of $\Psi_{qc}$ and
its complex conjugate coincide with vacuum averages of products of $\hat \Psi $ and $\hat \Psi^{\dagger}$:
\begin{equation}
<a_ka^{\ast}_{k^{'}}>=\delta^n(k-k^{'}), \quad <b_kb^\ast_{k^{'}}>=\delta^n(k-k^{'}),
\label{e41}
\end{equation}
where $<...>$ denotes a statistical average from now on and all other correlators are equal to zero.

Let us calculate the averaged square of the absolute value of $\Psi_{qc}$, ${\cal P} = <\Psi_{qc}\Psi^\ast_{qc}>$ giving an average probability
to find a universe. In the beginning we formally
assume that $\omega_L \sim \omega_{crit} $. Clearly, in this limit ${\cal P}_\ast\equiv {\cal P}(\omega_s=\omega_{crit}) =\bra 0 \hat \Psi {\hat \Psi}^{\dagger}\ket 0$. We have
\begin{equation}
{\cal P}_\ast={\pi^{-n/2}\over 2^{n}\Gamma (n/2)}\sqrt{{3\over \Lambda}}\,e^{-3t}\left ( 4e^{{2\over \Lambda }}I_1\sin^2 \phi +e^{-{2\over \Lambda}}I_2\cos^2 \phi \right ),
\label{e42}
\end{equation}
where
\begin{equation}
I_1=\int^{\infty}_0d\omega {\omega^{n-1}\over \sinh ({\pi \omega \over 2})}=2{2^n-1\over \pi^n}(n-1)!\zeta (n), \quad I_2=\int_0^{\omega_{crit}}d\omega
\omega^{n-1}\sinh \left({\pi \omega \over 2}\right)\sim {1\over \pi}\omega_{crit}^{n-1}e^{{\pi \omega_{crit} \over 2}},
\label{e43}
\end{equation}
where we use the same approximations as in the previous Section, $\zeta(n)$ is the Riemann zeta function and the integration limit
in the first integral is formally extended from $\omega_{crit}$ to infinity since the integrand there decreases exponentially with $\omega $.
Remembering that  $\omega_{crit}={2\over \sqrt 3}{1\over \Lambda}$ we see that the second term in (\ref{e42}) is much smaller than the first
and can be discarded. Note that the integrals in (\ref{e42}) are logarithmic diverging when $n=1$. We leave this special case for a future paper and assume from now on that we have more than one massless field in our model, $n \ge 2$. Neglecting the second term we can rewrite (\ref{e42})
in the form
\begin{equation}
{\cal P}_\ast=C_ne^{{2\over \Lambda}-3t}\sin^2 \phi, \quad C_n={(2^n-1)\over 2^{n-3}\pi^{{3\over 2}n}}{(n-1)!\zeta (n)\over \Gamma ({n\over 2})}\bigg ({3\over \Lambda }\bigg )^{1/2}.
\label{e44}
\end{equation}
Remarkably, the expression (\ref{e44}) is the same as the probability distribution obtained with help of the Hartle-Hawking wavefunction \cite{hh},
for a closed Universe with Lambda term, see e.g. \cite{kif}, his equation (8.63) with $V(\phi)={\Lambda \over 3}$.

Now we are going to estimate a correction accounting for $\omega_L$ being smaller than $\omega_{crit}$, but much larger than unity. In this case
the upper limit of integration in $I_1$ should be $\omega_L$. We have $I_1=\int^{\omega_{L}}_0d\omega {\omega^{n-1}\over \sinh ({\pi \omega \over 2})}
=\int^{\infty}_0d\omega {\omega^{n-1}\over \sinh ({\pi \omega \over 2})}-\Delta I$, where $\Delta I$ has the same integrand, but with integration
being performed from $\omega_s$ to $\omega_{crit}$. Approximating $\sinh (x)$ there as ${1\over 2}e^x$ we obtain
\begin{equation}
\Delta I \approx {2^{n+1}\over \pi^n}\left({\pi \omega_L\over 2}\right)^{n-1}e^{-{\pi \omega_L\over 2}}= {2^{n+1}\over \pi^n}\xi^{-1}_L\sqrt{\Lambda \over 3}
\left ({2\over \Lambda}-3t+\ln \left(\sqrt{{\Lambda \over 3}}\xi_L^{-1}\right)\right )^{n-1}e^{-({2\over \Lambda}-3t)},
\label{e45}
\end{equation}
and
\begin{equation}
{\cal P}={\cal P}_*(1-\Delta I/I_1)
\label{e46}
\end{equation}
Equations (\ref{e45}) and (\ref{e46}) tell that when $t \ll t_{crit}$ the correction determined by the dependency of $\omega_L$ on
time is exponentially small and the averaged quasi-classical wavefunction in our Universe is that of Hartle and Hawking.

Neglecting the contribution to (\ref{e42}) proportional to $I_2$ corresponds to discarding the term proportional to $\cos \phi $ in
(\ref{e24}). In this case equation (\ref{e40}) can be written as a product of a time dependent regular factor and a stochastic function
over the spatial variables
\begin{equation}
\Psi_{qc}={2\over \sqrt{\pi}}\left ({3\over \Lambda }\right)^{1/4}e^{{1\over \Lambda}-{3\over 2}t}\sin \phi
\int d^nk \left (C_{\omega}e^{ik\cdot x}a_k +C^\ast_{\omega}e^{-ik\cdot x}{b}_k^{\ast} \right )
\label{e47}
\end{equation}

In the end let us briefly discuss the opposite case $t \gg t_{crit}$. In this case almost all modes except those with very low frequencies
are in the regime of high frequency quasi-classical behaviour called 'case two' in the previous Section. In this regime, the modes momenta
are quasi-classical, and, accordingly, the time derivative of the wavefunction can be treated as a random classical variable, while
the wavefunction itself is essentially quantum. Differentiating (\ref{e47}) over the time we obtain an explicit expression for the time
derivative
\begin{equation}
\dot \Psi_{qc}={2\over \sqrt{\pi}}\left ({\Lambda \over 3 }\right)^{1/4}e^{{1\over \Lambda}+{3\over 2}t}\cos \phi
\int d^nk \left (C_{\omega}e^{ik\cdot x}a_k +C^\ast_{\omega}e^{-ik\cdot x}{b}_k^{\ast} \right )
\label{e48}
\end{equation}
Note that equation (\ref{e44}) tells that a probability to find a universe with the scale factor larger than critical is exponentially damped
on average. On the other hand from equation (\ref{e48}) it follows that the average of $\dot \Psi_{qc}$ always grows with time.

\subsection{Density matrix}
\label{wave2}
\begin{figure}[htbp]
	\begin{center}
		\includegraphics[width=0.6\textwidth]{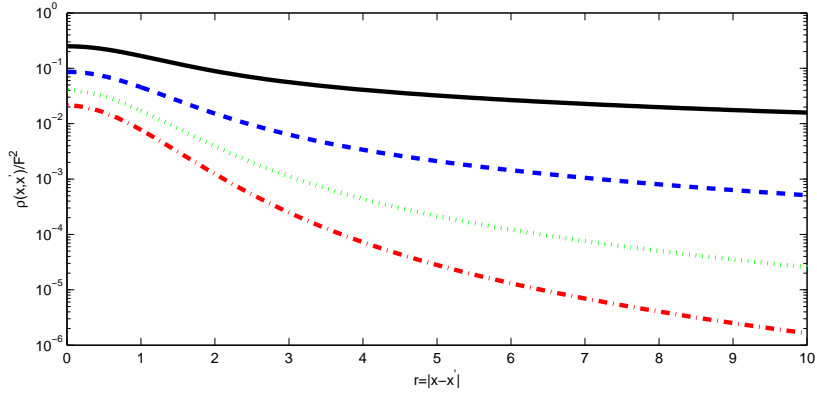} 
		\caption{We show the dependence of the ratio $\rho(x,x')/F^2$ as a function of the relative
distance $r=|x-x'|$. Solid, dashed, dotted and dot-dashed curves are calculated for $n=2$, $3$, $4$ and $5$, respectively.}
		\label{fig3}
	\end{center}
\end{figure}

Let us show that the stochastic wavefunction introduced above allows for an alternative description in terms of a density matrix with c-number valued matrix elements. For simplicity we assume below that we are in the regime when the wavefunction is approximately classical and, accordingly, $1\ll a \ll a_{crit}$. Also, we use sometimes hereafter the
bra and ket notation, for example representing (\ref{e47}) as $\Psi_{qc}=\bra x  \ket \Psi$. Let us stress that we are going to use both position representation associated with the field amplitudes, $\varphi_{i}$, as well as the adjoint momentum representation, which are clearly unrelated to the position and momentum representations considered in Section \ref{wave1}, where the dynamics of a mode amplitude of the wavefunction is treated.

In order to calculate explicitly the density matrix
$\rho (x, x')$ corresponding to (\ref{e47}) it is convenient to use temporarily the momentum representation taking into
account that $\bra k \ket x ={1\over (2\pi)^{n/2}}e^{-ikx}$. Making the Fourier transform of (\ref{e47}) we get an expression
similar to (\ref{e10}) with the exception that now the wavefunction in the momentum representation is considered as being
c-number valued:
\begin{equation}
\Psi_k\equiv \bra k \ket \Psi = (2\pi)^{n/2}F(a)(C_{\omega}a_{k}+C^{\ast}_{\omega}b_{-k}), \quad F(a)={2\over \sqrt{\pi}}\left ({3\over \Lambda }\right)^{1/4}e^{{1\over \Lambda}-{3\over 2}t}\sin \phi,
\label{e49}
\end{equation}
where we remind that $a_{k}$ and $b_{k}$ are classical complex Gaussian random numbers with correlation properties given by (\ref{e41}). It is convenient to represent these in terms of real random quantities as
\begin{equation}
a_{k}={1\over 2}\big (\alpha_1+\alpha_3+i(\alpha_4-\alpha_2)\big ),\quad b_{-k}={1\over 2}\big (\alpha_1-\alpha_3+i(\alpha_4+\alpha_2)\big ), \quad <\alpha_i\alpha_k>=\delta_{ik}\delta^{n}(k-k'),
\label{e50}
\end{equation}
introduce real and imaginary parts of $\Psi_k$ and $C_{\omega}$ as $R=Re(\Psi_k)$, $I=Im(\Psi_k)$, $A_{\omega}=Re(C_{\omega})$
and $B_{\omega}=Im(C_{\omega})$, and represent (\ref{e49}) in the form
\begin{equation}
R=(2\pi)^{n/2}F(a)(A_{\omega}\alpha_1+B_{\omega}\alpha_2), \quad I=(2\pi)^{n/2}F(a)(A_{\omega}\alpha_4+B_{\omega}\alpha_3).
\label{e51}
\end{equation}
Now, since both $R$ and $I$ are sums of two uncorrelated random Gaussian numbers the general theorem tells that
their distribution functions ${\cal P}(R)$ and  ${\cal P}(I)$ are also Gaussian, with the square of dispersion $\sigma^2=
<RR>=<II>$, while their joint distribution function ${\cal P}(R,I)= {\cal P}(R){\cal P}(I)$. Explicitly, we
have
\begin{equation}
{\cal P}(R,I)={1\over 2\pi \sigma^2}\exp{\left(-{|\Psi_k|^2\over 2\sigma^2}\right)}, \quad \sigma^2=(2\pi)^nF^2|C_{\omega}|^2.
\label{e52}
\end{equation}
The expression (\ref{e52}) gives a probability to find $\Psi_k$ with particular values of $R$ and $I$. Using this
fact the density matrix in the momentum representation, $\rho(k,k')$, can be obtained by the usual rule
$\rho(k,k')=\int dRdI \Psi_k\Psi^{\ast}_{k'}{\cal P}(R,I)$, where we can set $k=k'$ under the integral taking into account
that  $\rho(k,k')$ is clearly diagonal in this representation. In this way we obtain
\begin{equation}
\rho(k,k')=2(2\pi)^nF^2|C_{\omega}|^2\delta^{n}(k-k'), \quad |C_{\omega}|^2={\pi \over 4(2\pi)^n}\sinh^{-1}\left({\pi \omega \over 2}\right),
\label{e53}
\end{equation}
where we use (\ref{e18}). The expression (\ref{e53}) tells that mainly the modes of $\Psi_{qc}$ with small values of
$\omega$ give contributions to the mixed state defined by $\rho(k,k')$, while the contribution of modes with $\omega \gg
1$ is exponentially damped, see also the end of this Section.

The density matrix in the position representation, $\rho(x,x')={1\over (2\pi)^n}\int d^nkd^nk' e^{i(kx-k'x')}\rho(k,k')$ is easily obtained from
(\ref{e53}) taking into account that we can extend the upper limits of integrals over $k$ and $k'$ to infinity, since the
contribution of modes with $\omega > \omega_{s}$ to them is negligible provided that $\omega_{s} \gg 1$. We have
\begin{equation}
\rho(x,x')={F^{2}r^{-{n-2\over 2}}\over 2^{{n\over 2}+1}\pi^{{n\over 2}-1}}\int_{0}^{\infty}ds s^{{n\over 2}}J_{{n-2\over 2}}(rs)\sinh^{-1}\left({\pi \omega \over 2}\right),
\label{e54}
\end{equation}
where $r=|x-x'|$, $J_{\nu}(z)$ is the Bessel function and we use the standard result that a
multidimensional Fourier transform of a function depending only on the modulus of the momentum vector takes the form of one-dimensional  Hankel transform. Note that, since the Bessel function in (\ref{e54}) has a half-integer index, it can be expressed through elementary functions.

That the density matrix is a function only of the relative distance in the field space is clearly due to the homogeneity of our model with respect to the field coordinates, which, in its turn, is valid only for massless fields.
In general, for the potentials $V$ depending on the field coordinates the homogeneity is broken and the density matrix
must be determined by the field coordinates themselves. The dependence of $\rho(x,x')$ on $r$ is shown in Fig. \ref{fig3}
for several values of $n$.

The diagonal elements of the density matrix  $\rho(x, x)$ give probabilities to find fields with particular values of
$x$. They can be easily obtained from (\ref{e45}) using the decomposition of Bessel function near the origin of its
argument and the first expression in (\ref{e43}). In this way we obtain $\rho(x,x)={\cal P}_*$, where ${\cal P}_*$ is given by equation (\ref{e44}) as it should be.

\begin{figure}[htbp]
	\begin{center}
		\includegraphics[width=0.6\textwidth]{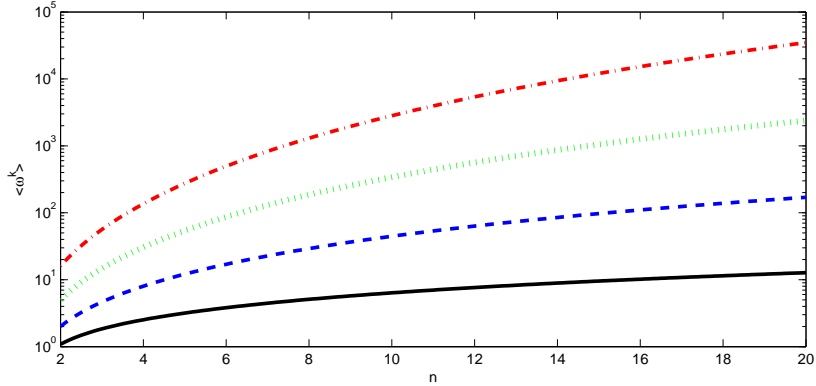} 
		\caption{The dependence of the expectation values given by equation (\ref{e56}) is shown as
a function of $n$. Solid, dashed, dotted and dot-dashed curves are calculated for $k=2$, $3$, $4$ and $5$, respectively.}
		\label{fig4}
	\end{center}
\end{figure}

It is well known that the Fourier transform of (\ref{e54}) determines a probability distribution of particles (universes,
in our case) over $k$, and, accordingly, over the field velocities $\dot \varphi_i$, see e.g. \cite{pit}. Then
equation (\ref{e53}) tells that it is proportional to $|C(\omega)|^2$ and we employ this result to obtain a normalised
probability distribution over $\omega $, ${\cal P}(\omega)$, as
\begin{equation}
{\cal P}(\omega)={\omega^{n-1}|C(\omega)|^2\over \int^{\infty}_{0} d\omega  \omega^{n-1}|C(\omega)|^2}=
{\omega^{n-1} \sinh^{-1} ({\pi \omega \over 2}) \over\int^{\infty}_{0} d\omega  \omega^{n-1} \sinh^{-1} ({\pi \omega \over 2})}=
{\pi^n \over 2(2^n-1)(n-1)!\zeta (n)} \omega^{n-1} \sinh^{-1} \left({\pi \omega \over 2}\right),
\label{e55}
\end{equation}
where we use (\ref{e43}) and (\ref{e53}). Let us remind that in the classical limit $\omega $ is proportional to
the absolute value of the 'total' field velocity, $\omega \propto v_{tot}=\sqrt{\sum_{i=1}^{n}\dot \varphi^2_i}$. Thus, equation (\ref{e55}) may be used to find probabilities of universes with different $v_{tot}$. This, in its
turn, can be employed to determine natural initial conditions for classical evolution of the Universe (or its sufficiently homogeneous parts, see Discussion below). It is instructive to calculate different expectation values
of powers of $\omega$
\begin{equation}
<\omega^k>={1\over \pi^k}{(n+k-1)!\over (n-1)!}{(2^{n+k}-1)\over (2^n-1)}{\zeta (n+k)\over \zeta (n)},
\label{e56}
\end{equation}
where we use again (\ref{e43}). The dependence of $<\omega^k>$ on $n$ is shown in Fig. \ref{fig4} for k=2..5, respectively.

\section{Conclusions and Discussion}
\label{conc}

We show, in the framework of a simple thirdly quantized minisuperspace model of a closed FRW universe with a small Lambda term and $n$ massless
scalar fields that its wavefunction operator has a simple interpretation in the limit of large scale factors $a \gg 1$ provided that
a natural $in$-vacuum state is specified for the system. Namely, when $1\ll a \ll e^{{2\over 3\Lambda }}$ the wavefunction operator may be
approximately treated as a classical random field with its averaged value being proportional to the Hartle Hawking wavefunction. When $a \gg e^{{2\over 3\Lambda }}$ it is the derivative of the wavefunction, which has the property of a classical random field
\footnote{Let us note that in certain minisuperspace models considering
classical WdW wave functions they also demonstrates 
quasiclassical (in this case WKBJ-like)
behaviour only at sufficiently small values of scale factors, see e.g. 
\cite{jul}.}.

The physical explanation of this result is the same as in the well developed theory of creation of a test field or density inhomogenities/gravitational
waves in a inflationary Universe. Both models have a copious production of excitations with respect to a suitably chosen $out$-vacuum state.
In both models the $in$-vacuum in the Schrodinger picture evolves to a strongly squeezed state, while the Wigner function of a particular mode
has a special form at large values of the natural time variable (or, the scale factor)
being proportional to a product of a Gaussian distribution of one of the
canonical variables with large dispersion and a distribution strongly peaked about a relation involving canonical variables, which is valid
on solutions to classical equations of motion.

Thus, the natural $in$-vacuum for the quantum WdW model provides a well defined classical although stochastic wavefunction at sufficiently large values of
the scale factor. The fact that its averaged value is proportional to the Hartle Hawking result needs a further investigation. Perhaps, some
progress could be made in the path integral formulation of our theory, although of course the classical action entering the Hartle Hawking
formalism should be substituted by the action induced by Lagrangian (\ref{e4}) in our case.

We also show that the stochastic wavefunction formalism is equivalent to the presence of a density matrix describing
a mixed state. Similar to the wavefunction approach, the matrix elements can be treated as c-numbers at sufficiently large
values of scale factors. Thus, in the framework of our model a mixed state defined over classical solutions of WdW equations emerges in this asymptotic limit. Diagonal elements of the density matrix giving a probability to find some particular values of fields and scale factor are again proportional to the corresponding expression following from the Hartle and Hawking formalism, thus the density matrix has a trivial form in the position representation. However, in the momentum representation
the dependence is non-trivial, it gives a probability to find universes with different values of field velocities. We
calculate this as well as associated expectation values in Section \ref{wave2}. It is important to stress that this result
is different from what follows from the Hartle-Hawking wavefunction, since the latter is uniform in the position representation
for the model with massless fields, and, therefore, predicts that field velocities are strictly zero in this case.

Clearly, that the density matrix gives a non-trivial distribution in the momentum representation is due to
the fact that in our models the field velocities have a direct physical meaning, while the field amplitudes are defined
up to arbitrary constant values. In principal, the distributions of this kind can be used to specify the most natural
initial conditions for classical evolution of the Universe.

Note, however, that program is hampered by the usual problems with interpretation of wavefunction and meaning of measurements
in quantum cosmology. Indeed, for example, it would be difficult to probe such distributions without a consideration of
an ensemble of universes with observers belonging to different universes able to communicate with each other. Of course,
it is difficult to achieve even when we take into account that the universes could 'interact' through quantum interference.
The difficulty can be alleviated either in models, where there is a non-linear interaction among the universes or
in a modification of the model, where it is assumed that it describes sufficiently large locally homogeneous parts of positive curvature of a generally inhomogeneous Universe. In the latter case
measurements can be performed in all parts and compared with each other after, say, the Lambda term decays due to
some reason and these parts come into causal contact with each other. Similar approaches were recently discussed in
e.g. \cite{bj} in the framework of loop quantum gravity and \cite{gc} in a thirdly quantized model.

It is interesting to point out that when the Universe is assumed to be slightly inhomogeneous, there is another mechanism of emergence of a mixed state by averaging out inhomogeneous degrees of freedom, see e.g. \cite{pad} or \cite{mor}. Also, let us note that the density matrix approach has been considered in the third quantization formalism, although in a sense quite different from what is discussed in this Paper, see e.g. \cite{tk}.

We suspect that a similar behaviour of wavefunction (or, emergence of the 'classical' density matrix) at large values of $a$ is present in all models, where a large production of excitations is observed.

Note that, strictly speaking, our results are valid for the number of massless fields $n \ge 2$. When $n=1$ the integrals $I_1$ and $I_2$
in equation (\ref{e43}) experience a logarithmic divergency at small values of $\omega $. This is also similar to the well known
logarithmic infrared divergency of test field in inflationary models, see e.g. \cite{lindeb} and references therein.
Although this case definitely needs an additional study, our formalism may be used to define
various conditional probabilities.

Finally, it would be very important to generalise our approach to the much more realistic case of a massive scalar field in a FRW universe.
In this case the separation of variables in the WdW equation is absent. However, this equation can be tackled numerically and our approach could be used to interpret numerical results. It is expected that the massive field case could have a non-trivial dependence of the density matrix on the field coordinates.

\begin{acknowledgements}
We are grateful to D. A. Kompaneets, A. A. Starobinsky and V. N. Strokov for valuable discussions, D. M. Gangardt and V. N. Lukash for useful remarks and Roma Basko for stimulating questions.
This study was supported in part by the grant of the President of the Russian Federation for Support of Leading
Scientific Schools NSh-4235.2014.2.
\end{acknowledgements}


\end{document}